\RequirePackage{lineno}

\documentclass[aps,pre,oldlfont,amsmath,amsfonts,amssymb,twocolumn]{revtex4}
\usepackage{bm}        % for math

\usepackage[latin1]{inputenc}
\usepackage{graphicx}
\usepackage[english]{babel}
\usepackage[usenames,dvipsnames]{color}

\usepackage{amsfonts}
\usepackage{amsmath}
\usepackage{amssymb}

\usepackage{color}

\usepackage{soul}
\newcommand{\beq}{\begin{equation}}
\newcommand{\eeq}{\end{equation}}
\newcommand{\nn}{\nonumber}
\newcommand{\ket}[1]{|#1\rangle}
\newcommand{\bra}[1]{\langle #1|}

\newcommand{\ra}{\rightarrow}

\makeatletter
\@ifundefined{textcolor}{}
{%
 \definecolor{BLACK}{gray}{0}
 \definecolor{WHITE}{gray}{1}
 \definecolor{RED}{rgb}{1,0,0}
 \definecolor{GREEN}{rgb}{0,1,0}
 \definecolor{BLUE}{rgb}{0,0,1}
 \definecolor{CYAN}{cmyk}{1,0,0,0}
 \definecolor{MAGENTA}{cmyk}{0,1,0,0}
 \definecolor{YELLOW}{cmyk}{0,0,1,0}
 }

\begin{document}

%\setpagewiselinenumbers
%\modulolinenumbers[5]
%\linenumbers

\title{Quantum thermophoresis}

\author{Maur\'icio Matos
$^{1}$
}
%\email{thiago.werlang80@gmail.com}
%ORCID: 0000-0001-5659-175X
%ResearcherID: E-7102-2012

\author{Thiago Werlang
$^{1}$
}
%\email{thiago.werlang80@gmail.com}
%ORCID: 0000-0001-5659-175X
%ResearcherID: E-7102-2012

\author{Daniel Valente
$^{1}$
}
\email{valente.daniel@gmail.com}
%ORCID: 0000-0002-3709-9118
%ResearcherID: I-9986-2018

\affiliation{
$^{1}$ 
Instituto de F\'isica, Universidade Federal de Mato Grosso, Cuiab\'a, MT, Brazil
}

\begin{abstract}
Thermophoresis is the migration of a particle due to a thermal gradient.
Here, we theoretically uncover the quantum version of thermophoresis.
As a proof of principle, we analytically find a thermophoretic force on a trapped quantum particle having three energy levels in $\Lambda$ configuration.
We then consider a model of $N$ sites, each coupled to its first neighbors and subjected to a local bath at a certain temperature, so as to show numerically how quantum thermophoresis behaves with increasing delocalization of the quantum particle.
We discuss how negative thermophoresis and the Dufour effect appear in the quantum regime.
\end{abstract}

%\pacs{03.65.Yz, 03.67.-a}

\maketitle
\begin{figure}[!htb]
\centering
\includegraphics[width=1.0\linewidth]{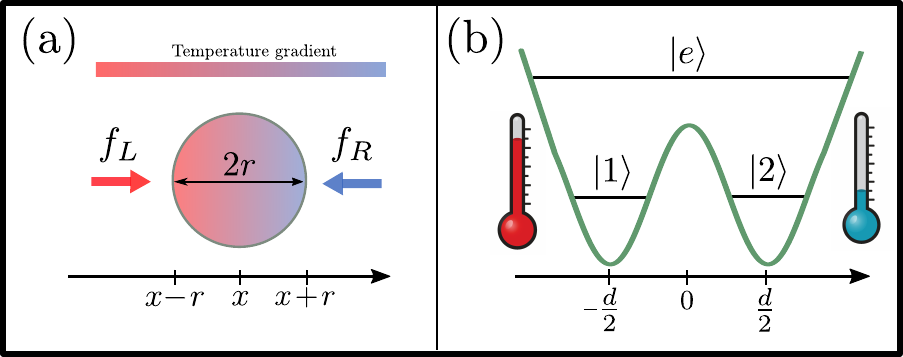} 
\caption{
(a) Classical thermophoresis in a particle of size $2r$ due to a thermal gradient.
The hotter (left) bath pushes with force $f_L$ the particle towards the coldest (right) bath, which pushes with force $f_R$ towards the other direction.
The net effect leads to Eq.(\ref{fclass}).
(b) Quantum thermophoresis in a $\Lambda$ three-level system.
A double-well potential illustrates the real-space representation of the system. 
The two minima are separated by $d$ (large enough so that no tunneling takes place; the model with tunneling is illustrated in Fig.(\ref{fig2})).
The higher temperature at the left bath 
(driving $\ket{1} \leftrightarrow \ket{e}$ transitions),
as compared to that at the right bath 
(driving $\ket{2} \leftrightarrow \ket{e}$ transitions),
makes jumps from $\ket{1}$ to $\ket{2}$ more likely than the reverse, so population concentrates closer to the cold bath.
We derive a quantum thermophoretic force for this model
(see Eqs.(\ref{qT}) and (\ref{qTclass})).
}
\label{fig1}
\end{figure}

%%%%%%%%%%%%%%%%%%%%%%%%%%%%%%%%%%%%%%%
%\section{Introduction}
%%%%
A particle can move from hot to cold, an effect known as thermophoresis \cite{matsuo2000,braun}.
This can be understood in terms of an asymmetrical Langevin force on a Brownian particle.
Let us consider a one-dimensional model where the position $x$ of a Brownian particle of length $2r$ fluctuates due to a Langevin force given by $f(x) = f_L(x-r) - f_R(x+r)$, which arises from a gas of microscopic particles (see Fig.\ref{fig1} (a)).
Terms $f_{L,R}(x)$ tell apart left-side from right-side collisions with the environment particles.
If each side is at a slightly different temperature, and considering that each local force can be written as $f_{L,R}(x) = \xi(t) \sqrt{T(x \mp r)}$, where $T(x)$ is a spatial-dependent temperature \cite{matsuo2000}, we can write the thermophoretic force as a finite average Langevin force,
\beq
\langle f(x) \rangle = - \frac{T'(x)}{T(x)} \ p V,
\label{fclass}
\eeq 
up to first order in $r$.
Here, $p$ is the local pressure, and $V \propto r^3$ is an effective volume.
The important point in Eq.(\ref{fclass}) is that it reveals how a negative gradient can imply a positive force, pushing the particle from hot to cold.

Thermophoresis has proven relevant in diverse contexts. 
It has been spotlighted as a key mechanism for exponential acceleration of RNA polymerization \cite{dieter1}, with possible implications to the origin of life on earth \cite{dieter2,busiello}.
It has also been identified in the stochastic dynamics of antiferromagnetic solitons \cite{kim}.
However, thermophoresis has so far remained restricted to the domain of classical motion.
Could it also emerge from a quantum mechanical formulation, as hypothesized in Ref.\cite{cp22}?

Here, we show that thermophoresis can take place in elementary quantum systems.
We start with a quantum particle having three energy levels in $\Lambda$ configuration, coupled to a pair of independent bosonic baths at different temperatures (see Fig.\ref{fig1} (b)).
When the particle is at state $\ket{1}$ (which can be interpreted as a localized state around position $-d/2$ in a bistable potential), it interacts with a hot bath which promotes a transition to the excited state $\ket{e}$.
If the particle spontaneously decay towards state $\ket{2}$, it can get trapped there (i.e., around position $d/2$), as the cold bath may not provide enough thermal energy for the particle to reach $\ket{e}$ again (tunneling is not allowed in this model, which is a reasonable assumption as long as $d$ is sufficiently large).
On average, this creates an accumulation of population in the ground state coupled to the cold bath.
In this sense, the quantum thermophoresis we present here can be seen as a hot-to-cold migration in Hilbert space.
However, we also wish to find a quantum force in real space, as analogous to Eq.(\ref{fclass}).
%%%

By writing down the equation of motion for an average position that we associate to this quantum particle, we derive a quantum thermophoretic force given by
\beq
F_q = - \frac{\delta n}{2} \ m^* \Gamma^2 d ,
\label{qT}
\eeq
where the main thermal effect comes from 
$\delta n \equiv n_2 - n_1$,
in terms of the bosonic average numbers of excitations,
$n_k = [\exp(\hbar \omega_k/k_B T_k)-1]^{-1}$.
Here, $\omega_k = (E_e-E_k)/\hbar$ are the transition frequencies, and $T_k$ are the bath temperatures, with $k_B$ being the Boltzmann constant.
The effective mass of the particle, $m^*$, also depends on the temperatures,
$m^* = 1/(4\bar{n} + 2)$, 
where
$\bar{n} \equiv (n_2 + n_1)/2$.
The couplings of the system to its quantized baths give rise to the spontaneous emission rates $\Gamma$.

In the high-temperature (semiclassical) limit, we find that
\beq
F_q^{\mathrm{(high \ T)}} \approx - \frac{T'}{T} \frac{\Gamma^2 d^2}{6},
\label{qTclass}
\eeq
which reminds us of Eq.(\ref{fclass}) mainly due to the 
$-T'/T$ 
term, where 
$T' \equiv (T_2-T_1)/d$, 
and 
$T \equiv (T_1+T_2)/2$.
Beyond the thermal gradient, Eqs.(\ref{fclass}) and (\ref{qTclass}) also bear resemblance in the fact that $\Gamma^2$ is an environment-dependent term, much like the pressure $p$ in the classical model, and $d$ gives the system's length scale, as does $r$ classically.

\begin{figure}[!htb]
\centering
\includegraphics[width=1.0\linewidth]{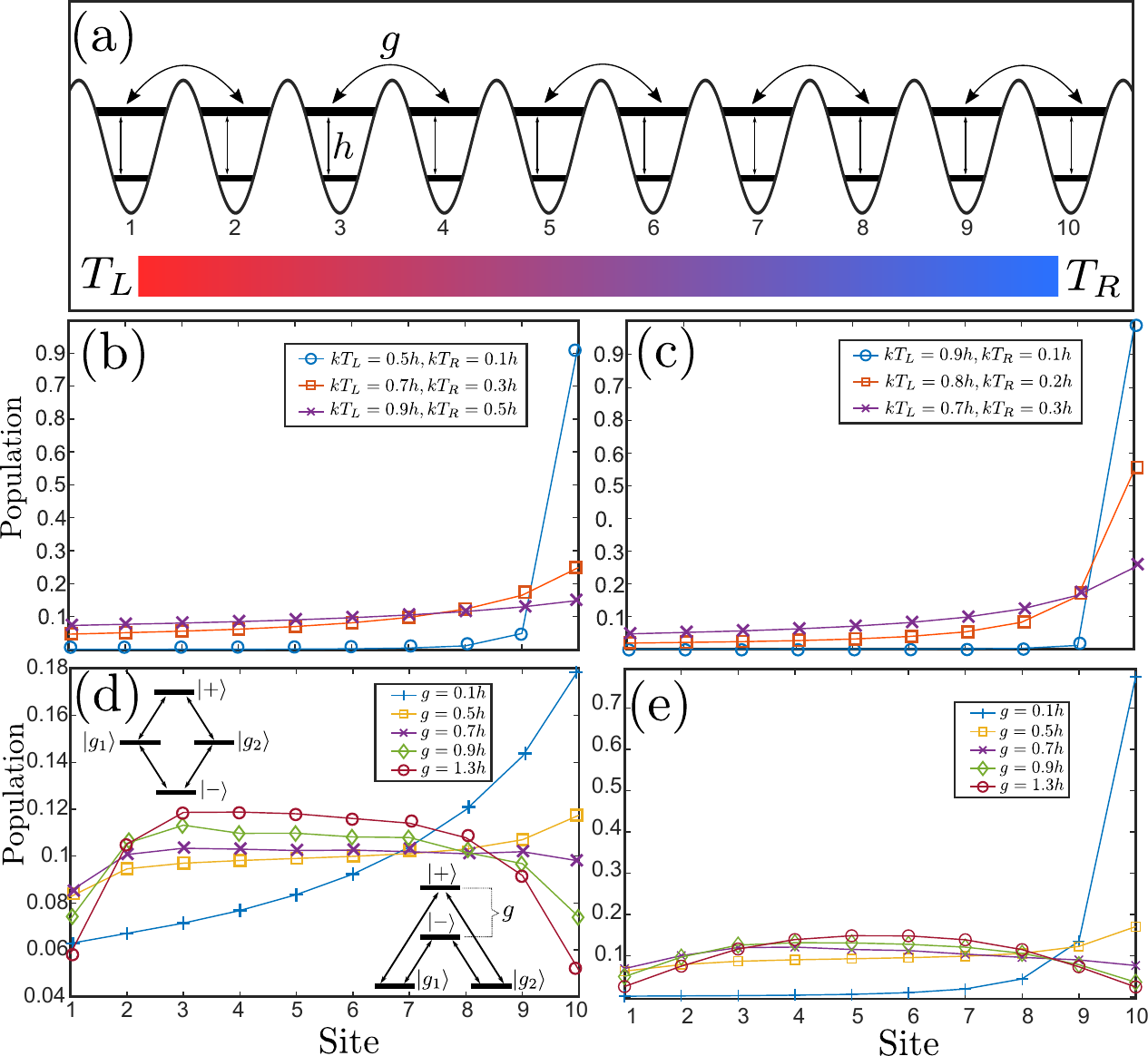} 
\caption{
(a) $N$-site model (here, we take $N=10$).
Each site has a local energy $h$ and coupling $g$ to its first neighbors, and is locally coupled to an environment at a given temperature, linearly varying from $T_L$ to $T_R$.
(b) and (c) show quantum thermophoresis for a fixed gradient (b), and fixed average temperature (c), both at $g=0.1h$.
(d) and (e) show how either negative thermophoresis or full delocalization can take place, depending on the coupling strength $g$, as well as on the temperatures.
In panel (d), $k_BT_L=0.8h$ and $k_BT_R=0.4h$. 
In panel (e), $k_BT_L=0.3h$ and $k_BT_R=0.1h$.
In all the plots, the population of the $i$-th site (from $1$ to $10$) is the sum of its local ground plus its excited-state populations, in the steady-state regime.
}
\label{fig2}
\end{figure}

To go a step further, we investigate how quantum thermophoresis behaves for a more delocalized quantum particle.
We numerically analyze a one-dimensional (1D) lattice containing $N=10$ two-level sites, each coupled to its first neighbors by means of a quantum coherent tunneling rate $g$, and subjected to a local temperature.
Strength $g$ allows us to control the degree of delocalization of the quantum particle in our model.
See Fig.\ref{fig2}(a).
We compute the steady-state population of each site (sum of the ground and excited state populations) at a given $g$, and temperatures $T_{L,R}$ (we assume a linear gradient, here).
Our results, shown in Figs.\ref{fig2}(b) to (e), are explained below in more detail.
In short, they evidence either positive or negative thermophoresis (i.e., migration towards the hot region \cite{prl19}), depending on $g$ and $T_{L,R}$.
We also analytically demonstrate that negative thermophoresis shows up in a three-level model, but in $V$ configuration.

To conclude the paper, we discuss how the so called Dufour effect \cite{onsager}, namely, an induction of a thermal gradient due to a fixed nonequilibrium particle concentration (the reciprocal of thermophoresis), also holds in the quantum regime.
%%%

%%%%%%%%%%%%%%%%%%%%%%%%%
{\it Formalism.--}
In all the models, we use a system-plus-reservoir approach, where the total hamiltonian is
$H = H_S + H_I + H_E$.
The system is generically described by
$H_S = \sum_n E_n \ket{n}\bra{n}$.
The environment Hamiltonian is given by $N$ sets of independent harmonic oscillators, namely,
$H_E = \sum_{k=1}^N H_E^{(k)}$, so that
$H_E^{(k)} = \sum_j \hbar \omega_j a_{k,j}^{\dagger} a_{k,j}$, forming a continuum of frequencies $\omega_j$.
The interactions with the baths are given by 
$H_I = \sum_{k=1}^N H_{I}^{(k)}$, with
$H_I^{(k)} = \sigma_x^{(k)} \sum_j \hbar g_{k,j} (a_{k,j} + a_{k,j}^\dagger)$.
We have defined $\sigma_x^{(k)}$ as the system's local degree of freedom that is affected by the $k-$th environment.
The continuum limit is described by the spectral function
$J^{(k)}(\omega) = 2\pi \sum_j g_{k,j}^2 \delta(\omega - \omega_j)$.
In the models we analyze here, we assume that
$J^{(k)}(\omega) = \Gamma_k$ (i.e., frequency independent).
In the derivation of Eqs.(\ref{qT}) and (\ref{qTclass}), we have used that $\Gamma_1 = \Gamma_2 = \Gamma$ for simplicity.

We derive a Markovian quantum master equation using the so called microscopic approach \cite{petruccione,werlang,werlang15,prl16,pre20,cp22,pereira}, where we consider the eigenstates of $H_S$ to form the appropriate basis onto which the thermal baths act.
This is particularly relevant to the 1D lattice model illustrated in Fig.\ref{fig2}(a), where a phenomenological (local) approach may lead to quite different outcomes \cite{werlang,pre20}.
In terms of the reduced density operator of the system $\rho_S$, our master equation reads
\beq
\partial_t \rho_S(t)=-(i/\hbar)[H_S,\rho_S(t)]+L[\rho_S(t)],
\label{ME}
\eeq
where $L[\rho_S(t)] = \sum_{k=1}^N L_k$ describes the thermal effects, where \cite{petruccione,werlang15}
\begin{align}\label{L}
L_k = \sum_{\omega} 
\gamma_k(\omega)
\Big[ 
A^{(k)}_\omega \rho_S  A^{(k)\dagger}_\omega
-\frac{1}{2}\left\{ \rho_S, A^{(k)\dagger}_\omega A^{(k)}_\omega \right\}
\Big].
\end{align}
Here, $\omega=\omega_{ij}=(E_j-E_i)/\hbar$, with $\gamma_k(\omega) = J^{(k)}(\omega) (1+n_k(\omega))$ for $\omega>0$.
For $\omega<0$, we have that $\gamma_k(\omega) = J^{(k)}(|\omega|) n_k(|\omega|)$.
Again, $n_k(\omega)$ are the bosonic thermal averages.
The jump operators are defined by
$A^{(k)}_\omega = \sum_{i,j|\omega=\omega_{ij}} \ket{i}\bra{i} \sigma_x^{(k)}\ket{j}\bra{j}$.

As a final remark, we mention that, if $\omega = 0$, we have that 
$\gamma_k(0) = \lim_{\omega\rightarrow0} (J^{(k)}(\omega)/\hbar\omega) k_B T_k$,
which converges for the ohmic case, $J^{(k)}(\omega) = \eta_k \omega$, but diverges for the constant spectral functions that we are using here throughout the entire paper.
In our models, all the operators $A^{(k)}_{\omega = 0}$ vanish, so we have no divergence in our results.
We have also tested with ohmic spectral densities (not shown), finding qualitatively identical results to those presented in Fig.(\ref{fig2}).

%%%%%%%%%%%%%%%%%
{\it Lambda system.--}
In our three-level model, the excited state $\ket{e}$ is coupled to the lowest eigenenergies $\ket{1}$ and $\ket{2}$ (see Fig.\ref{fig1}(b)).
That is, the number of thermal baths is $N=2$, and the coupling operators are 
$\sigma_x^{(k)} = \ket{e}\bra{k} + \ket{k}\bra{e}$, for $k=1,2$.
In this case, we find from Eq.(\ref{ME}) that the populations $P_n(t) = \bra{n} \rho_S(t) \ket{n}$ obey to the rate equations
\beq
\partial_t P_k = \Gamma (n_k + 1) P_e - \Gamma n_k P_k.
\label{rateeq}
\eeq
We eliminate the excited state population with the help of probability conservation, $P_e = 1 - P_1 - P_2$.

We recast the populations in terms of the unbalance
\beq
\Delta \equiv P_2 - P_1,
\eeq
along with the average ground-state population 
$P \equiv (P_2+P_1)/2$.
In the steady-state, we find that
\beq
\Delta_{ss} = - \frac{\delta n}{\bar{n}(3\bar{n}+2) - (3/4) \delta n^2},
\label{deltass}
\eeq
showing how the particle tends towards the state coupled to the cold bath.
In the extreme case where $T_2 = 0$ (hence $n_2 = 0$), we have that $\bar{n} = -\delta n /2$, so that
$\Delta_{ss} = 1$, which means full concentration at the cold side.

Eq.(\ref{rateeq}) clearly shows that the quantum nature of the bath is key to thermophoresis in the present model:
If we replace 
$n_k + 1 \rightarrow n_k$, 
which amounts to ignoring the spontaneous emission term arising from quantum fluctuations in the environment, we get that 
$\Delta_{ss} = 0$, 
hence no thermophoresis regardless of the thermal gradient.
This is equivalent to taking the $\bar{n} \rightarrow \infty$ limit for a fixed $\delta n$ in Eq.(\ref{deltass}).

To go beyond thermophoresis in the Hilbert space, so as to show its connection with real-space thermophoresis, we map our $\Lambda$ system into a bistable potential where $\ket{1}$ and $\ket{2}$ represent two localized minima separated by distance $d$ sufficiently large so that no tunneling is allowed (see Fig.\ref{fig1}(b)).
We consider the average position as given by
\beq
\langle X \rangle = -\frac{d}{2} P_1 + \frac{d}{2} P_2 = \frac{d}{2} \ \Delta,
\eeq
where we have assumed that the excited state is symmetrical in the sense that 
$\bra{e} X \ket{e} = 0$.
From Eqs.(\ref{rateeq}), we find that
\beq
m^* \langle \ddot{X} \rangle 
+ \Gamma \langle \dot{X} \rangle
+ m^* \Omega^2 \langle X \rangle 
= 
- ( m^* \Gamma^2 d ) \ \delta n  / 2,
\label{qx}
\eeq
which we interpret as a driven-damped harmonic oscillator with effective mass 
$m^* = 1/(4\bar{n}+2)$, 
damping rate $\Gamma$,
and frequency $\Omega$, as given by
$\Omega^2 \equiv \Gamma^2 \left[\bar{n} (3\bar{n} + 2) - (3/4) \delta n^2 \right]$.
Note that $\Omega^2 \geq 0$, since $n_{1,2} \geq 0$, so the gradient is bounded to $|\delta n| \leq 2 \bar{n}$.
The RHS of Eq.(\ref{qx}) shows the thermophoretic nature of the driving force, $F_q \propto - \delta n$, as we have shown in Eq.(\ref{qT}).

We highlight Eq.(\ref{qx}) as our main result.
The stationary unbalance of the populations in the lowest levels of a $\Lambda$ atom, as shown in Eq.(\ref{deltass}), is by itself not surprising.
Indeed, this very effect has been used in Ref.\cite{walmsley} as an experimental resource for achieving a quantum advantage in a thermal machine, given the population inversion it produces when the cold side has smaller transition frequency.
By contrast, our work reveals that Eq.(\ref{deltass}), along with its application in Ref.\cite{walmsley}, correspond to special cases of a much broader nonequilibrium phenomena known as thermophoresis, now encompassing its quantum and classical counterparts.

We can now discuss the high-temperature limit of our three-level model, which we define as
$n_k \approx k_B T_k/\hbar \omega_k \gg 1$. 
In this limit, we have analytically verified that 
$\langle \ddot{X} \rangle \approx - \Gamma \bar{n} \langle \dot{X} \rangle$, 
which we substitute in Eq.(\ref{qx}) so as to obtain an equation of motion akin to the overdamped regime of the oscillator, namely,
\beq
\Gamma \langle \dot{X} \rangle 
+ \frac{4m^* \Omega^2}{3} \langle X \rangle 
=
-  \frac{\delta n}{3 \bar{n}} \frac{\Gamma^2 d}{2}.
\label{overdampedqx}
\eeq
We now interpret the RHS of Eq.(\ref{overdampedqx}) as the high-temperature limit of the thermophoretic force, namely,
\begin{align}
F_q^{\mathrm{(high T)}} 
\equiv 
-  \frac{\delta n}{3 \bar{n}} \frac{\Gamma^2 d}{2} 
&\approx 
- \frac{T' d/\hbar \omega}{3 T/\hbar \omega} \frac{\Gamma^2 d}{2} \nn\\
&= -\frac{T'}{T} \frac{\Gamma^2 d^2}{6}.
\label{deriveqTclass}
\end{align}
To unambiguously distinguish thermophoresis from the thermal equilibrium bias, we have set $\omega_1 = \omega_2 = \omega$ in Eq.(\ref{deriveqTclass}).

%%%%%%%%%%%%%%%%%%%%%%%%
{\it N-site model.--}
The $\Lambda$ model reveals quantum thermophoresis for a somewhat localized quantum particle.
This contrasts with the classical free Brownian motion used as our starting point.
However, a free quantum particle has a continuous spectrum, so our master equation cannot be applied under such a condition.
The quantum theory of Brownian motion (the so called Caldeira-Leggett model \cite{caldeira}) is far more suitable to treat a continuous spectrum, but is nevertheless restricted to a thermal-equilibrium environment.

To address this issue, we consider a 1D model consisting of $N$ sites.
Each site has two energy levels separated by $h$, with the excited state of each site $\ket{e_k}$ coupled to its neighbors by means of a quantum tunneling rate $g$, so the Hamiltonian reads
$
H_S = \sum_{k=1}^{N} h \ket{e_k}\bra{e_k} + \sum_{k=1}^{N-1} g (\ket{e_k} \bra{e_{k+1}} + \mbox{h.c.}).
$
See Fig.(\ref{fig2})(a).
This type of model has been extensively employed in tight-binding approaches, including studies on photosynthetic complexes where a relatively small number of sites is typically considered ($N \leq 7$ in Refs.\cite{njp2008,njp2010}, for instance).
Here, the key difference is that we assume that each site is locally coupled to an environment at a given (possibly site-dependent) temperature.
Because the spectrum is kept discrete, we have no problem in employing our master equation, as given in Eqs.(\ref{ME}) and (\ref{L}).

Our numerical results are shown in Fig.(\ref{fig2}), for $N=10$.
Thermophoresis appears in panels (b) and (c), both showing that the closer to the cold bath the site is, the more populated it gets.
In (b), we fix the gradient, $T_L-T_R$, while varying the average temperature, $(T_L+T_R)/2$.
In (c), we vary the gradient, keeping the average temperature fixed.
In both cases, we assume small couplings, $g = 0.1h$.
Thermophoresis is of course more pronounced in the regimes of lowest average temperatures, and highest gradients.

In Fig.(\ref{fig2})(d) and (e), we explore particle delocalization by increasing $g$.
At large couplings, $g/h \gtrsim 1$, the ground state becomes delocalized, in the sense that the ground state of the system does not correspond to the ground state of the sites
(see the level crossing as depicted in panel (d) for $N=2$, where the eigenstates are 
$\ket{\pm} = (\ket{e_1}\pm\ket{e_2})/\sqrt{2}$ and $\ket{g_{1,2}}$).
We find two distinct behaviors at large couplings, depending on the temperatures.
At lower temperatures, the particle becomes symmetrically delocalized among the sites for $g=1.3h$ 
(the pick gets centered around sites $5$ and $6$).
See panel (e), where we set $T_L = 0.3 h/k_B$ and $T_R = 0.1h/k_B$.
Delocalization in panel (e) is not flat because sites $1$ and $10$ form closed boundaries.
Instead, the particle is more concentrated at the middle, resembling the ground state of a quantum particle in a box.
At higher temperatures, an anomaly becomes quite evident.
See panel (d), where we set $T_L = 0.8h/k_B$ and $T_R = 0.4h/k_B$.
At $g=1.3h$, site $3$ concentrates the highest population, which decreases as going from $4$ to $10$.
This means a migration towards the hottest bath rather than the coldest one, unravelling the quantum version of an effect known as negative thermophoresis \cite{prl19}.
We further examine that point below.

%%%%%%%%%%%%%%%%%%%%%%%%
{\it Negative quantum thermophoresis.--}
To better understand how a negative thermophoresis appears in the $N$-site model, as numerically shown above, we discuss a single three-level system in $V$ configuration, where the same effect can be analytically demonstrated.
Now, the energy levels are such that $E_1 \sim E_2 \gg E_g$.
We also associate the $V$ system with a particle in real space, in close analogy with the $\Lambda$ system.
That is, we assume that
$\langle X \rangle_V \equiv (d/2) [P_2 - P_1]$
can be considered as the average position of a certain quantum particle, where $P_{1,2}$ are the populations of two excited states of the $V$ system.
We thus find that
\beq
m^*\langle \ddot{X} \rangle_V
+ \Gamma \langle \dot{X} \rangle_V
+ m^* \Omega_v^2 \langle X \rangle_V
=
F_V,
\eeq
where 
$\Omega_V^2 = \Gamma^2 \left[ (3\bar{n} + 1)(\bar{n} + 1) - (3/4) \delta n^2 \right]$,
and
\beq
F_V \equiv + \frac{\delta n}{2} \ m^* \Gamma^2 d = - F_q.
\label{negth}
\eeq
Eq.(\ref{negth}) reveals migration of a quantum particle towards the hot bath, therefore characterizing a negative quantum thermophoresis.
Remarkably, it also shows that $F_V$ is the exact negative of the $\Lambda$ thermophoretic force $F_q$ (see Eqs.(\ref{qT}) and (\ref{qx})).

Going back to the $N$-site model, we see that large couplings $g\gtrsim h$ effectively create a $V$-type structure.
This explains why in Fig.(\ref{fig2})(d) for $g = 1.3h$ the population concentrates at site $3$, closer to the hottest bath, and decreases from site $4$ to $10$ towards the coldest bath.
Eq.(\ref{negth}) also clarifies how negative thermophoresis nonlinearly depends on the ratio between the temperatures and the gaps, through $\delta n$, helping us to understand why it was much less visible in Fig.(\ref{fig2})(e) 
(we do see a tiny bit of negative thermophoresis in panel (e), for $g/h=0.7$, and $0.9$, but far less pronounced than in (d) at $g/h=1.3$).
This is so because $\delta n \ra 0$ if the baths are both too cold with respect to a fixed energy gap, making the populations of both the excited states $\ket{1}$ and $\ket{2}$ of the $V$ system to vanish, wiping out the negative thermophoresis.

%%%%%%%%%%%%%%%%%%%%%%%%
{\it Quantum Dufour effect.--}
The Dufour effect is the onset of a thermal gradient due to a given particle concentration gradient.
This is reciprocal to thermophoresis, as described by the Onsager relations \cite{onsager}.
We look for analogous phenomena in the quantum regime.
We find that, by assuming a heterogeneous population distribution in the $\Lambda$ system, the heat dissipated from the system to its thermal equilibrium environment can be unbalanced.
If the environment has finite heat capacity, the temperature rises unevenly.
We call this is a quantum Dufour effect.

In order to see this, we define, as in Ref.\cite{cp22}, the heat current from the system to the baths as
$J_{k} = - \mbox{Tr}[\mathcal{L}_{k}(\rho_S) H_S]$,
for $k=1,2$, in the case of the $\Lambda$ system.
We get that
$J_{k} = \hbar \omega_k \Gamma \ [(n_{k} + 1) P_e - n_{k} P_{k}]$.
We again set $\omega_1 = \omega_2 = \omega$.
First, we depart from thermal equilibrium, $T_1 = T_2$, which implies that $n_1 = n_2 = n$.
Now, let us assume a given heterogeneous population distribution, say, $P_2 > P_1$ 
(with population inversion, $P_e > P_k \ n/(n+1)$, so as to guarantee positive heat currents).
In this case, we find that $J_1 > J_2 > 0$, which means that the heat is dissipated faster towards bath $1$ coupled with the transition $\ket{1} \leftrightarrow \ket{e}$ than towards bath $2$, coupled with the $\ket{2} \leftrightarrow \ket{e}$ transition.
If the heat baths have finite and equal heat capacities, $T_1$ will raise faster than $T_2$, thus creating a thermal gradient.
This confirms the presence of a quantum Dufour effect in the $\Lambda$ system.

%%%%%%%%%%%%%%%%%%%%%%%%
{\it Discussions.--}
In summary, we have unraveled the phenomenon of thermophoresis in the ultimate quantum regime.
The three-level $\Lambda$ quantum system is arguably the most elementary scenario where quantum thermophoresis may take place, where a thermophoretic force has been shown to obey $F_q \propto - \delta n$ in the low-temperatures regime, and $\propto - T'/T$ in the high-temperature semiclassical limit.
The quantum nature of the environment plays a key role in quantum thermophoresis, as we have discussed below Eq.(\ref{deltass}).
With the help of a 1D model with $N$ coupled sites, we have studied how quantum thermophoresis survives for a more delocalized quantum particle.
We have also uncovered a negative thermophoretic effect, which we have analytically explained with a three-level system in $V$ configuration, showing a thermophoretic force 
$F_V = - F_q$.
We point out that the quantum negative thermophoresis here is for weak system-bath couplings, in contrast to the strong-coupling regime attributed to the negative thermophoresis in Ref.\cite{prl19}.
Finally, we have shown that the complementary effect of thermophoresis, a.k.a. the Dufour effect, also manifests itself in the quantum regime.

As a remark, we notice that the term ``quantum thermophoresis'' has first appeared in Ref.\cite{ss}, as far as we know, where the authors have theoretically studied the classical motion of a mesoscopic particle swimming in a mixture of superfluid and normal fluid phases.
Generalizing quantum thermophoresis to more diverse models will contribute to a broader understanding of nonequilibrium self-organization due to thermal gradients, to which our work represents a first step.

\begin{acknowledgements}
Instituto Nacional de Ci\^encia e Tecnologia de Informa\c c\~ao Qu\^antica (465469/2014-0). 
Conselho Nacional de Desenvolvimento Cient\'ifico e Tecnol\'ogico (402074/2023-8).
M. M. was supported by CAPES.
\end{acknowledgements}


\begin{thebibliography}{10}

\bibitem{matsuo2000} Miki Matsuo, and Shin-ichi Sasa, 
Stochastic energetics of non-uniform temperature systems,
Physica A \textbf{276}, 188 (2000).

\bibitem{braun} Stefan Duhr, and Dieter Braun,
Why molecules move along a temperature gradient,
Proc. Natl. Acad. Sci. USA \textbf{103}, 19678 (2006).

\bibitem{dieter1} Christof B. Mast, Severin Schink, Ulrich Gerland, and Dieter Braun,
Escalation of polymerization in a thermal gradient,
Proc. Natl. Acad. Sci. USA \textbf{110}, 8030 (2012).

\bibitem{dieter2} Annalena Salditt, Lorenz M. R. Keil, David P. Horning, Christof B. Mast, Gerald F. Joyce and Dieter Braun,
Thermal Habitat for RNA Amplification and Accumulation,
Phys. Rev. Lett. \textbf{125}, 048104 (2020).

\bibitem{busiello} Daniel Maria Busiello, Shiling Liang, Francesco Piazza, and Paolo De Los Rios,
Dissipation-driven selection of states in non-equilibrium chemical networks,
Comm. Chem. \textbf{4}, 16 (2021).

\bibitem{kim} Se Kwon Kim, Oleg Tchernyshyov, and Yaroslav Tserkovnyak,
Thermophoresis of an antiferromagnetic soliton,
Phys. Rev. B \textbf{92}, 020402(R) (2015).

\bibitem{cp22} Thiago Werlang, Maur\'icio Matos, Frederico Brito, and Daniel Valente,
Emergence of energy-avoiding and energy-seeking behaviors in nonequilibrium dissipative quantum systems,
Comm. Phys. \textbf{5}, 7 (2022).

\bibitem{prl19} Rodrigo de Miguel, and J. Miguel Rub\'i,
Negative Thermophoretic Force in the Strong Coupling Regime,
Phys. Rev. Lett. \textbf{123}, 200602 (2019).

\bibitem{onsager} Lars Onsager, 
Reciprocal Relations in Irreversible Processes. I.,
Phys. Rev. \textbf{37}, 405 (1931).

\bibitem{petruccione} H. P. Breuer and F. Petrucionne, 
The Theory of Open Quantum Systems, 
Oxford University Press (2002).

\bibitem{werlang} T. Werlang, M. A. Marchiori, M. F. Cornelio and D. Valente,
Optimal rectification in the ultrastrong coupling regime,
Phys. Rev. E {\bf 89}, 062109 (2014).

\bibitem{werlang15} T. Werlang and D. Valente, 
Heat transport between two pure-dephasing reservoirs,
Phys. Rev. E {\bf 91}, 012143 (2015).

\bibitem{prl16} Karl Joulain, J\'er\'emie Drevillon, Youn\`es Ezzahri, and Jose Ordonez-Miranda,
Quantum Thermal Transistor,
Phys. Rev. Lett. \textbf{116}, 200601 (2016).

\bibitem{pereira}
Emmanuel Pereira,
Perfect thermal rectification in a many-body quantum Ising model,
Phys. Rev. E \textbf{99}, 032116 (2019).

\bibitem{pre20} D. Valente and T. Werlang,
Frustration and inhomogeneous environments in relaxation of open chains with Ising-type interactions,
Phys. Rev. E \textbf{102}, 022114 (2020).

\bibitem{walmsley} James Klatzow, Jonas N. Becker, Patrick M. Ledingham, Christian Weinzetl, Krzysztof T. Kaczmarek, Dylan J. Saunders, Joshua Nunn, Ian A. Walmsley, Raam Uzdin, and Eilon Poem,
Experimental Demonstration of Quantum Effects in the Operation of Microscopic Heat Engines,
Phys. Rev. Lett. \textbf{122}, 110601 (2019).

\bibitem{caldeira} A. O. Caldeira,
An Introduction to Macroscopic Quantum Phenomena and Quantum Dissipation,
Cambridge University Press (2014).

\bibitem{njp2008} M. B. Plenio and S. F. Huelga,
Dephasing-assisted transport: quantum networks and biomolecules,
New J. Phys. \textbf{10}, 113019 (2008).

\bibitem{njp2010} Stephan Hoyer, Mohan Sarovar and K. Birgitta Whaley,
Limits of quantum speedup in photosynthetic light harvesting,
New J. Phys. \textbf{12}, 065041 (2010).

\bibitem{ss} G. V. Kolmakov, and I. S. Aranson,
Superfluid swimmers,
Phys. Rev. Res. \textbf{3}, 013188 (2021).


%\bibitem{JCP2013} England, J. L.
%Statistical physics of self-replication.
%J. Chem. Phys. {\bf 139,} 121923 (2013).
%
%\bibitem{petra17} Schwille, P. 
%How Simple Could Life Be?
%{\it Angew. Chem. Int. Ed.}
%{\bf 56,} 10998-11002 (2017).
%
%\bibitem{nnano2020} te Brinke, E., et al. 
%%Joost Groen1, Andreas Herrmann2,3, Hans A. Heus   1, GermÃ¡n Rivas4, Evan Spruijt   1* and Wilhelm T. S. Huck1*
%Dissipative adaptation in driven self-assembly leading to self-dividing fibrils.
%{\it Nat. Nanotech.}
%{\bf 13,} 849-855 (2020).
%
%\bibitem{nnano2015} England, J. 
%Dissipative adaptation in driven self-assembly.
%{\it Nat. Nanotech.}
%{\bf 10,} 919-923 (2015).
%
%\bibitem{goldenfeld06} Vetsigian, K., Woese, C., and Goldenfeld, N. 
%Collective evolution and the genetic code. 
%{\it Proc. Natl. Acad. Sci. USA}
%{\bf 103,} 10696-10701 (2006).
%
%\bibitem{prx16} Perunov, N., Marsland, R. A., and England, J. L.
%Statistical physics of adaptation.
%{\it Phys. Rev. X}
%{\bf 6,} 021036 (2016).
%
%\bibitem{nphys12} Lan, G., Sartori, P., Neumann, S., Sourjik, V., and Tu, Y.
%The energy--speed--accuracy trade-off in sensory adaptation.
%{\it Nat. Phys.}
%{\bf 8,} 422-428 (2012).
%
%\bibitem{PRE2015} Kondepudi, D., Kay, B., and Dixon, J.
%End-directed evolution and the emergence of energy-seeking behavior in a complex system.
%{\it Phys. Rev. E}
%{\bf 91,} 050902(R) (2015).
%
%\bibitem{huck16} Grzybowski, B. A., and Huck, W. T. S.
%The nanotechnology of life-inspired systems.
%{\it Nat. Nanotech.}
%{\bf 11,} 585-592 (2016).
%
%\bibitem{ncomm17} Ilday, S. et al.
%Rich complex behaviour of self-assembled nanoparticles far from equilibrium.
%{\it Nat. Comm.}
%{\bf 8,} 14942 (2017)
%% doi: 10.1038/ncomms14942
%
%\bibitem{PRL2017} Kachman, T., Owen,  J. A., and England, J. L.
%Self-organized resonance during search of a diverse chemical space.
%{\it Phys. Rev. Lett.}
%{\bf 119,} 038001 (2017).
%
%\bibitem{PNAS2017} Horowitz, J. M. and England, J. L.
%Spontaneous fine-tuning to environment in many-species chemical reaction networks.
%{\it Proc. Natl. Acad. Sci. USA}
%{\bf 114,} 7565 (2017).
%
%\bibitem{nmat2017} Bachelard, N. et al.
%%Chad Ropp, Marc Dubois, Rongkuo Zhao, Yuan Wang and Xiang Zhang
%Emergence of an enslaved phononic bandgap in a non-equilibrium pseudo-crystal.
%{\it Nat. Mat.}
%{\bf 16,} 808-814 (2017).
%
%\bibitem{ragazzon18} Ragazzon G., and Prins, L. J.
%Energy consumption in chemical fuel-driven self-assembly.
%{\it Nat. Nanotech.}
%{\bf 13,} 882-889 (2018). 
%
%\bibitem{nphoton2018} Ropp, C., Bachelard, N., Barth, D., Wang, Y., and Zhang, X. 
%Dissipative self-organization in optical space.
%{\it Nat. Photon.}
%{\bf 12,} 739-743 (2018).
%
%\bibitem{kedia19} 
%Kedia, H., Pan, D., Slotine, J.-J., and England, J. L.
%Drive-specific adaptation in disordered mechanical networks of bistable springs.
%Preprint at: https://arxiv.org/abs/1908.09332 (2019).
%
%\bibitem{nphys2020} Makey, G., et al.
%%ÃzgÃŒn Yavuz  K?vanÃ§ GÃŒngÃ¶r 1,2, Sezin Galioglu   1, Roujin Ghaffari1, E. Doruk Engin   3, GÃ¶khan Y?ld?r?m   4,5, Onurcan Bekta?   6,7, Ã. Seleme Nizam   8, Ãzge Akbulut   9, ÃzgÃŒr ?ahin   9,10, 1, Didem Dede   1, H. Volkan Demir   1,2,5,11, F. Ãmer Ilday   1,2,5 and Serim Ilday
%Universality of dissipative self-assembly from quantum dots to human cells.
%{\it Nat. Phys.}
%{\bf 16,} 795-801 (2020).
%%https://doi.org/10.1038/s41567-020-0879-8
%
%\bibitem{cp21} Valente, D., Brito, F., Werlang, T.
%Quantum dissipative adaptation.
%{\it Comm. Phys.}
%{\bf 4,} 11 (2021).
%
%\bibitem{gt12} Krammer, H., M\"oller, F. M., and Braun, D.
%Thermal, autonomous replicator made from transfer RNA.
%{\it Phys. Rev. Lett.}
%{\bf 108,} 238104 (2012).
%
%\bibitem{gt15} Kreysing, M., Keil, L., Lanzmich S., and Braun, D.
%Heat flux across an open pore enables the continuous replication and selection of oligonucleotides towards increasing length.
%{\it Nat. Chem.}
%{\bf 7,} 203-208 (2015).
%
%\bibitem{gt17} Keil, L. M. R. {\it et al.}
%%Friederike M. Möller1, Michael Kieß1, Patrick W. Kudella1 & Christof B. Mast1
%Proton gradients and pH oscillations emerge from heat flow at the microscale.
%{\it Nat. Comm.}
%{\bf 8,} 1897 (2017).
%
%\bibitem{gt20} Salditt A. {\it et al.}
%%Lorenz M. R. Keil,1,* David P. Horning ,2,* Christof B. Mast,1 Gerald F. Joyce ,2 and Dieter Braun 1,?
%Thermal habitat for RNA amplification and accumulation.
%{\it Phys. Rev. Lett.}
%{\bf 125,} 048104 (2020).
%
%\bibitem{gt21} Busiello, D. M., Liang, S., Piazza F., and De Los Rios, P.
%Dissipation-driven selection of states in non-equilibrium chemical networks.
%{\it Comm. Chem.}
%{\bf 4,} 16 (2021).
%
%\bibitem{gtsc} Chr\'etien, D. {\it et al.}
%%Paule Be? nit1,2, Hyung-Ho Ha3, Susanne Keipert4, Riyad El- Khoury5, Young-Tae Chang6, Martin Jastroch4, Howard T. Jacobs7,8, Pierre Rustin1,2,9*, Malgorzata Rak
%Mitochondria are physiologically maintained at close to 50 $^\circ$C.
%{\it PLoS Biol.}
%{\bf 16,} 1 (2018).
%
%\bibitem{w14} Werlang, T., Marchiori, M. A., Cornelio, M. F., and Valente, D.
%Optimal rectification in the ultrastrong coupling regime.
%{\it Phys. Rev. E}
%{\bf 89,} 062109 (2014).
%
%\bibitem{qc2} Werlang, T. and Valente, D.
%Heat transport between two pure-dephasing reservoirs.
%{\it Phys. Rev. E}
%{\bf 91,} 012143 (2015).
%
%\bibitem{qc3} Zhang, Z. and Wang, J.
%Landscape, kinetics, paths and statistics of curl flux, coherence, entanglement and energy transfer in non-equilibrium quantum systems.
%{\it New J. Phys}
%{\bf 17,} 043053 (2015).
%
%\bibitem{qc4} Fang, X., Kruse, K., Lu, T., and Wang, J.
%Nonequilibrium physics in biology.
%{\it Rev. Mod. Phys.}
%{\bf 91,} 045004 (2019).
%
%\bibitem{jcp20} Cook, J. and Endres, R. G.
%Thermodynamics of switching in multistable non-equilibrium systems.
%{\it J. Chem. Phys.}
%{\bf 152,} 054108 (2020).
%
%\bibitem{crooks} Crooks, G. E.
%Entropy production fluctuation theorem and the nonequilibrium work relation for free energy differences.
%{\it Phys. Rev. E}
%{\bf 60,} 2721 (1999).
%
%\bibitem{alicki} Alicki, R.
%The quantum open system as a model of the heat engine.
%{\it J. Phys. A: Math. Gen.}
%{\bf 12,} L103 (1979).
%
%\bibitem{prb} Wijesekara, R. T., Gunapala, S. D., Stockman, M. I., and Premaratne, M.
%Optically controlled quantum thermal gate.
%{\it Phys. Rev. B}
%{\bf 101,} 245402 (2020).


\end{thebibliography}
\end{document}